# Automated Standardization of Legacy Biomedical Metadata Using an Ontology-Constrained LLM Agent

**Josef Hardi, MSc[1], Martin J. O'Connor, MSc[1], Marcos Martínez-Romero, PhD[1], Jean G. Rosario, PhD[2], Stephen A. Fisher, PhD[2], and Mark A. Musen, MD, PhD[1]**

**Affiliations**
[1] Division of Computational Medicine, Stanford University, 3180 Porter Dr, Palo Alto, CA 94304, USA.
[2] Department of Biology, University of Pennsylvania, 433 S University Avenue, Philadelphia, PA 19104, USA.

**Corresponding author.** Josef Hardi, Division of Computational Medicine, Stanford University, 3180 Porter Dr, Palo Alto, CA 94304, USA. Email: johardi@stanford.edu
**Author email addresses.** sunid@stanford.edu (M.J. O'Connor), marcosmr@stanford.edu (M. Martínez-Romero), jrosar@sas.upenn.edu (J.G. Rosario), safisher@sas.upenn.edu (S.A. Fisher), musen@stanford.edu (M.A. Musen)

## Abstract

***Objective:*** *Descriptive scientific metadata in public repositories are often incomplete and inconsistent with community standards and ontologies, limiting data FAIRness. Large language models (LLMs) offer a promising approach to automatically standardizing such metadata when the LLM is provided with knowledge of the relevant standards in machine-actionable form (e.g., as metadata templates from systems such as the CEDAR Workbench). Entry of metadata standards through prompt engineering offers only fixed snapshots of the relevant standards, however, and it relies on an LLM's pretrained knowledge to interpret and satisfy constraints related to the standard. We evaluate whether giving an LLM access to metadata specifications and authoritative terminology at runtime improves automated metadata standardization.*

***Methods:*** *We present ARMS, a tool-using LLM agent that retrieves complete CEDAR metadata templates and that queries authoritative biomedical terminology services dynamically at execution time. We compared ARMS with a prompt-engineering approach on 839 legacy metadata records describing experiment datasets from the Human BioMolecular Atlas Program (HuBMAP) with expert-standardized records serving as the reference standard.*

***Results:*** *Across the corpus, ARMS outperformed the prompt-engineering approach in converting legacy metadata to standards-compliant form, increasing precision from 0.56 to 0.93 and recall from 0.51 to 0.85, with improvements across all field categories and assay types. The largest gains occurred for ontology-constrained fields, where precision increased from 0.36 to 0.92.*

***Conclusion:*** *Our findings show that access to appropriate metadata standards can substantially improve LLM-based standardization of legacy metadata. ARMS improves metadata standardization by giving LLMs runtime access to the authoritative resources that define valid metadata. Representation of metadata standards in machine-actionable form enhances the task of rectifying legacy metadata using LLM technology, especially when those standards can be queried dynamically.*

## 1. Introduction

Scientific research generates vast amounts of data that have value well beyond the original study—for experiment validation, secondary exploration, meta-analysis, and the training of computational models. Realizing that value requires that data be accompanied by high-quality, descriptive metadata: structured descriptions of the experimental conditions, samples, and methods that produced the data. The FAIR Guiding Principles articulate this requirement precisely, specifying that descriptive metadata must be rich, machine-interpretable, and adherent to community standards if data are to be Findable, Accessible, Interoperable, and Reusable [1]. In practice, descriptive metadata in public repositories are frequently incomplete, inconsistent, and poorly structured. Field names are idiosyncratic, field values are given as free text (including typos), and the listed attributes of the experiment are difficult to predict (and thus to search) [2]. The result is a large archive of research data that are difficult or impossible to discover and integrate.

Scientific communities have long recognized this problem and have responded by developing metadata reporting guidelines and standards that enumerate the descriptive information required for particular types of experimental data. The Minimum Information About a Microarray Experiment (MIAME), for example, specifies the types of fields required to describe a functional genomics experiment in sufficient detail to search and reuse the data [3]. Hundreds of such minimal information checklists now exist across the life sciences. However, these standards are commonly expressed as textual documents: They identify which fields should be present but provide less guidance regarding how values should be represented. An investigator who records a tissue sample as "lung tissue," "lung," or "pulmonary" is following the spirit of a standard while producing a value ambiguous to any computational system attempting to aggregate or query metadata across studies. The looseness of textual metadata standards is not a minor inconvenience; it is a fundamental barrier to making datasets FAIR and "AI ready" [4].

A practical way to operationalize a community-based metadata standard is to represent it as a reusable, machine-readable template. Such a template serves as a schema for metadata records: It defines which fields a record should contain, describes what those fields mean, and specifies rules governing their values. The Center for Expanded Data Annotation and Retrieval (CEDAR) Workbench, developed by our team, supports the creation, maintenance, and application of such templates [5–7]. In CEDAR, metadata are represented as a list of attribute–value pairs, and the CEDAR template that structures the metadata indicates, for each attribute, the datatype of the associated value. A template field may accept free text, a term from a predefined list of values, terms from a naming authority such as ORCID or RRID, or terms drawn from a designated ontology or ontology branch. Because these requirements are encoded as machine-readable constraints, software can use them to create, validate, and standardize metadata records.

A major challenge for data repositories and data coordinating centers is the sheer volume of legacy metadata created before tools such as CEDAR became available. Millions of records across public repositories may require retrospective standardization before their associated datasets can be reliably discovered, integrated, and reused [8]. Manual curation of these records is costly and difficult to scale, while rule-based transformations are often brittle because field names, value formats, and terminology vary widely among records. Large language models (LLMs) may help address this gap: They can interpret heterogeneous field names and values, use contextual information to infer correspondences, and transform records to match a target schema. Our prior research showed that LLMs such as GPT-4 when used by themselves are not very good at rectifying legacy metadata, but that the addition of appropriate CEDAR templates can guide GPT-4 in standardizing metadata to improve downstream data retrieval [9,10].

This work sits within a growing literature on LLMs for biomedical metadata standardization and related semantic tasks. LLMs may perform well on biomedical entity normalization, where free-text values are linked to controlled vocabulary terms, but performance degrades for terms that are infrequent in training data or that require selection from large ontological hierarchies [11,12]. Schema-guided approaches provide additional structure for this process. SPIRES, for example, uses LinkML schemas to guide the extraction of structured entities and relations from text and grounds the resulting entities against ontologies through external lookup, improving identifier accuracy over unassisted LLMs [13]. However, SPIRES addresses schema-guided extraction from text, whereas our task begins with existing metadata records that must satisfy the field-level constraints of a target standard. A key similarity is that both the LinkML schemas used by SPIRES and the CEDAR templates used in our task can restrict fields to a particular ontology, ontology branch, or value list, providing constraints that guide the LLM in standardizing the field values.

One common way to provide an LLM with external knowledge at inference time is retrieval-augmented generation (RAG), in which information retrieved from an external source is added to the model's context [14]. Conventional RAG systems typically retrieve relevant text passages, however, whereas metadata standardization requires structured operations such as restricting candidate values to a specified ontology or ontology branch, and verifying that the output satisfies a field-level constraint. LLMs with access to external tools can instead query services or databases as they reason [15], and have shown advantages on tasks requiring current or structured knowledge, including database queries, code execution, and factual verification [16,17]. Recent work on agentic data retrieval likewise shows that agents operating over structured semantic metadata retrieve machine-actionable datasets more precisely than do agents searching the unstructured Web, reinforcing that LLM reasoning alone does not obviate the need for explicit machine-readable structure [18].

We developed Agentic Real-Time Metadata Standardization (ARMS), a tool-augmented LLM system that transforms a legacy metadata record into a target schema by consulting machine-readable field constraints and authoritative biomedical terminologies at runtime. ARMS treats CEDAR field–value constraints as executable specifications enforced at runtime. It retrieves the live CEDAR template and queries BioPortal, a public repository of more than 1,200 biomedical ontologies [19], to identify and verify standard terms, including terms restricted to specified ontology branches. Rather than assuming that the LLM happens to store the ontology values that satisfy a constraint, ARMS gives the LLM access to the external resources that define those values.

We evaluate ARMS using a unique resource from the Human BioMolecular Atlas Program (HuBMAP), an NIH-funded consortium constructing a spatial atlas of the human body at single-cell resolution [20]. HuBMAP generates data across genomic sequencing, tissue imaging, and mass spectrometry, with assay-specific metadata specifications encoded as CEDAR templates [21]. Importantly, domain experts have manually standardized legacy HuBMAP metadata records against these specifications. These curated records provide a reference standard that is rarely available in metadata research, allowing us to evaluate standardized field values directly using exact matching rather than relying on proxy measures of metadata quality.

### Statement of Significance

| | |
|---|---|
| **Problem or Issue** | Legacy biomedical metadata rarely satisfy the value constraints that community standards require, and manual metadata curation does not scale to repository-sized archives. |
| **What is Already Known** | LLMs prompted with field names and ontology constraints can rectify legacy metadata records, but these constraints are provided only as plain text, requiring the model to rely on its training knowledge to interpret and satisfy them. |
| **What This Paper Adds** | An agent that retrieves machine-actionable template specifications and that queries terminology services at runtime achieves considerably higher precision and recall when compared with our previous prompt-engineering approach. Encoding metadata standards in a machine-actionable form not only can enhance metadata standardization prospectively, but also enables the conversion of legacy metadata to standard form at scale. |
| **Who would benefit from the new knowledge in this paper** | Developers who build automated pipelines for retrospective metadata standardization and anyone with an interest in making existing datasets "AI ready." |

## 2. Methods

*2.1 Tool-using agent*

We implemented ARMS in Python, wrapping an LLM with access to predefined tools for retrieving information from external knowledge sources. We use the Model Context Protocol (MCP) [22], an open standard that enables LLMs to interact with external services through a standardized server interface.

An ARMS execution begins with a user prompt containing two components: a legacy metadata record and a CEDAR template identifier. The agent is additionally governed by a system prompt that defines the transformation procedure, tool-use rules, and output requirements (Section 2.2). During execution, ARMS uses the CEDAR template identifier to retrieve the target template specification and uses the available BioPortal tools when values must be resolved against controlled terminologies in real time. The agent returns two outputs: the transformed metadata record, structured according to the target CEDAR template, and a processing log documenting the transformation. The rules governing how ARMS maps legacy fields to template fields and resolves their values are defined by the agent system prompt and described in the following section. Figure 1 illustrates the ARMS system architecture.

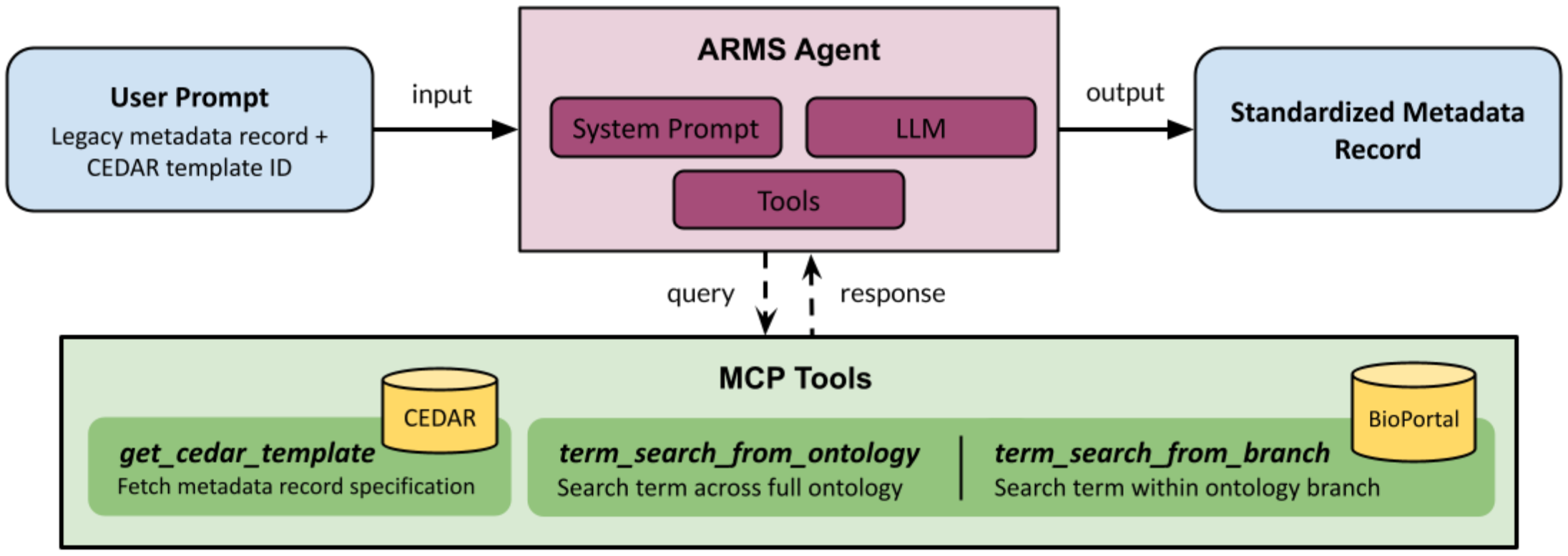

**Figure 1. Architecture of Agentic Real-Time Metadata Standardization (ARMS).** The agent receives a legacy metadata record and a target CEDAR template. Guided by the system prompt, it retrieves the complete template specification through the *get_cedar_template* tool exposed using the Model Context Protocol (MCP) and performs field mapping. When a field requires an ontology-constrained value, the agent uses *term_search_from_ontology* or *term_search_from_branch* tools to query BioPortal for candidate terms from the specified ontology or ontology branch. For non-ontology-constrained fields, the agent uses the retrieved field specification to look for a corresponding legacy field–value and to resolve the type and formatting. Finally, the agent outputs a standardized record.

### *2.2 Agent system prompt*

The ARMS agent's transformation behavior is governed by a single *system prompt*. Whereas the *user prompt* provides the record to be transformed and identifies its target metadata template, the system prompt specifies how the agent must perform that transformation generally. The system prompt is procedural rather than descriptive: Instead of describing what good metadata look like, it defines what the agent must determine, when it may call the available tools, and what it must return. The prompt contains rules governing several aspects of the agent's behavior. Here, we focus on the two that directly determine how values from a legacy metadata record are transformed into a target CEDAR template: field mapping and value resolution.

**Field Mapping.** The mapping process proceeds from the fields in the CEDAR metadata template to the fields in the legacy record. The agent considers each template field in turn and looks at the legacy record for the field or fields that contain the information that might populate it. The agent evaluates candidate matches using the field name and description in the template together with the names and values of the legacy fields, rather than relying on name similarity alone. A template field may therefore be mapped to a legacy field with a different name when the meanings and values indicate that the legacy field and the template field correspond. For example, ARMS may map a template field named `library_output_amount_value` to the legacy field `library_final_yield` if the field description, name, and value fit together. After this initial pass, the agent performs a second pass in the reverse direction. It takes each unused legacy field and considers whether it corresponds to any template field that remains unmapped, allowing it to identify matches that may have been overlooked in the first pass. Any legacy field that still cannot be placed is recorded in the processing log rather than forced into an unsupported template field.

**Value Resolution.** After a legacy field has been mapped to a template field**,** the standardization of the legacy field's value is guided by the constraints specified for that field in the CEDAR template. A field whose value is bound to an ontology or ontology branch requires a search tool call, using the legacy value as the search string, to find the closest valid term. A field constrained to a list of string literals is resolved by selecting from the enumerated values provided by the template. An unconstrained field keeps its legacy value verbatim, subject only to datatype and string pattern normalization. The agent may use its domain knowledge to interpret what is already present in the legacy record, but it may not make new assertions, supply missing details, or alter the original legacy values. An empty legacy value is treated as *unknown* rather than inapplicable. When no valid value can be determined, the agent leaves the field empty, because asserting a value that violates the template specification is worse than leaving the field unresolved.

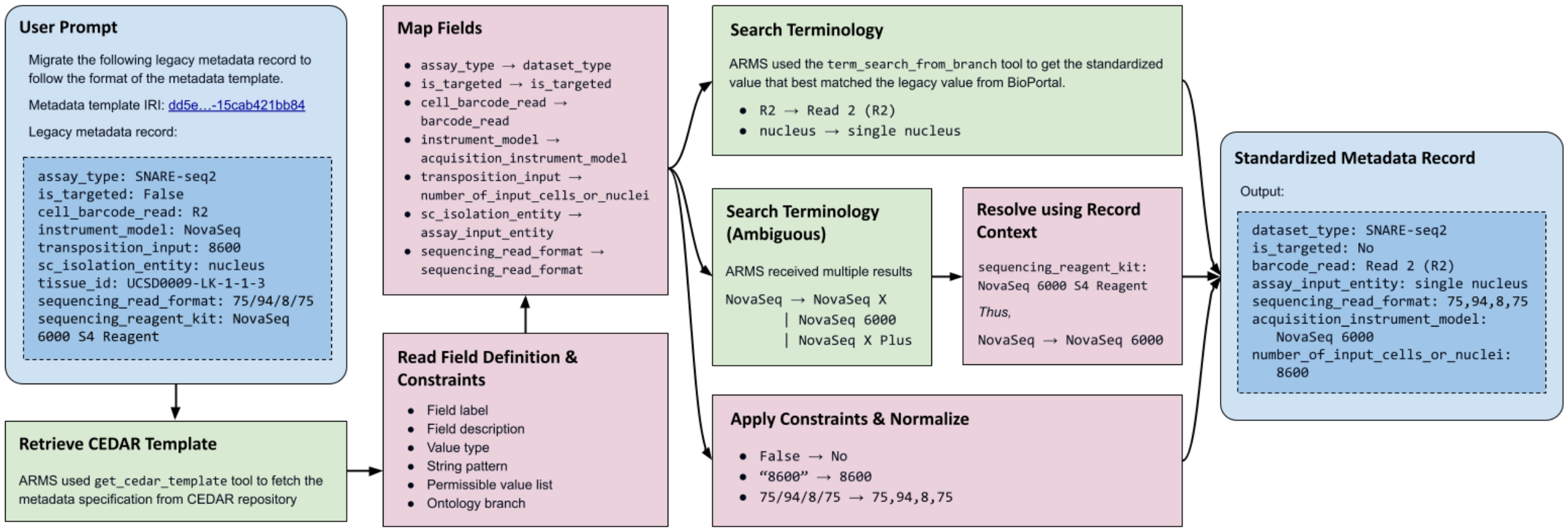

**Figure 2. Example ARMS metadata transformation workflow.** ARMS retrieves the target CEDAR template, maps legacy fields to template fields, and resolves their values according to the field constraints. Values constrained by controlled terminologies are resolved through BioPortal searches, with existing record context used to disambiguate candidate terms when necessary. Other values are selected from permissible values or normalized to satisfy datatype and string-pattern constraints. The resulting record conforms to the target CEDAR template while preserving the information mapped to template fields.

### *2.3 Example ARMS workflow*

Figure 2 illustrates how the field-mapping and value-resolution rules operate together for a single legacy metadata record. Given the legacy record and target CEDAR template identifier in the user prompt, ARMS first retrieves the template and reads each field's definition and constraints. It then maps legacy fields to their corresponding template fields. For fields constrained by controlled terminologies, ARMS searches the specified terminology branch for a standardized value that best matches the legacy value. When a terminology search returns multiple plausible results, ARMS uses information already present elsewhere in the record to resolve the ambiguity. In the example, the legacy value "NovaSeq" produces several candidate terms, and the model identifies "NovaSeq 6000" as the appropriate instrument model based on the `sequencing_reagent_kit` value. Finally, ARMS applies the template's permissible values, datatypes, and string patterns to normalize values to produce a record that conforms to the target template.

### *2.4 Prompt-only baseline*

We used the static, prompt-engineering approach from our prior work [10] as the baseline for measuring the value added by our dynamic, agentic methods. In our earlier work, an LLM rectifies legacy metadata using a target specification supplied as plain text in the prompt. To assess the contribution of ARMS's access to external knowledge sources, we implemented this baseline without the CEDAR and BioPortal tools. Its system prompt was derived from the ARMS system prompt and followed the same general procedures for field mapping and value resolution. For each legacy record, the user prompt supplied the target field names and the names of the vocabularies associated with ontology-constrained fields. The baseline approach mapped legacy fields to these target fields and resolved ontology-constrained values using the LLM's pretrained knowledge rather than by querying an authoritative ontology.

### *2.5 Evaluation dataset*

We obtained our reference dataset from 2,568 expert-curated legacy HuBMAP metadata records across 12 assay types. The original legacy records predated HuBMAP's adoption of CEDAR-based metadata specifications and were subsequently standardized against assay-specific CEDAR templates. An experimental biologist with expertise in histology, spatial proteomics, and RNA sequencing standardized and curated the records by producing field mappings, by correcting misplaced values and non-standard representations, and by verifying values against experimental protocols and manufacturer documentation. As a curation policy, legacy values containing useful partial information or references were preserved rather than discarded, even when they could not be fully standardized to the target specification. Similarly, when source documentation conflicted with the recorded metadata or provided insufficient information to resolve it, the original value was retained and flagged for further review. The resulting expert-curated records serve as the expert-curated reference standard for our evaluation.

To construct the evaluation set, we sampled the HuBMAP datasets independently within each assay type to ensure representation across all 12 assays. Each evaluation example consisted of a pair comprising an original legacy metadata record and its corresponding expert-standardized metadata record, with the legacy record serving as the input and the standardized record as the expected output. We selected up to 100 records uniformly at random per assay using a fixed random seed; when fewer than 100 records were available, all records were included. This process yielded 839 records: ATACseq (n=100), RNAseq (n=100), Auto-fluorescence (n=100), CODEX (n=100), MIBI (n=100), Histology (n=77), Cell DIVE (n=32), IMC-2D (n=13), Lightsheet (n=9), LC-MS (n=100), MALDI (n=93), and DESI (n=15).

The 839 standardized records contained 28,257 field instances (mean 33.7 per record). Assay-specific templates contained 23–48 fields, including 6–21 ontology-constrained fields. Because some values necessarily recur across records of the same assay type (for example, all RNAseq records share the same analyte class "RNA"), the 28,257 field instances correspond to 1,875 distinct assay-specific field–value pairs: 1,686 populated pairs (200 ontology-constrained and 1,486 non-ontology-constrained) and 189 blank pairs. A detailed breakdown of the evaluation dataset by assay is provided in Supplementary Material A (Table A.1).

*2.6 Evaluation metrics*

We compared each method's output with the reference standard field by field. Exact, case-sensitive matching was used because text fields in the HuBMAP metadata specifications define the required representation of values. Ontology-constrained fields and fields with permissible values require values to match the specified labels, while other text fields may define patterns that values must follow. The only exception was for DOI URLs, where doi.org and dx.doi.org were treated as equivalent. For ontology-constrained fields, the scored values were label strings rather than ontology identifiers or URIs. Non-string values, such as numbers, were compared by equality. Null values, empty or whitespace-only strings, and absent output fields were treated as missing. Thus, the metrics measure agreement with the expert-curated reference records.

We classified each field-level comparison into four outcomes. A **true positive** occurred when both records populated the field and the reference-standard and asserted values matched. A **false positive** occurred when the method asserted a value for a field that was blank in the reference-standard record (an insertion), whereas a **false negative** occurred when the method left a field blank for which the reference-standard record specified a value. When the method asserted a value that differed from the reference-standard value (a substitution), the comparison was counted as **both a false positive and a false negative**. From these outcomes, we computed precision and recall as our primary measures. Precision measures the proportion of values asserted by the method that match the reference standard, whereas recall measures the proportion of reference-standard values that the method successfully reproduces.

**Field categories.** We report precision and recall for two complementary field categories:

1. **Ontology-constrained fields** are fields whose permissible values are defined by an ontology or ontology branch.
2. **Non-ontology-constrained fields** are fields whose expected values are specified directly in the template, such as enumerations, datatypes, or patterns.

**Units of analysis.** Because the same assay-specific field–value pair can recur across multiple records, a single systematic correction may be counted many times. We therefore evaluate performance using two complementary units of analysis. *Field-level scoring* treats each field in each record as a separate unit, reflecting the correction workload: how many individual values are correctly rectified across the entire corpus. *Deduplicated scoring* reduces the influence of recurring values through averaging. For recall, we calculated the proportion of occurrences of each distinct populated assay-specific reference field–value pairs that were correctly reproduced, then averaged these proportions across distinct pairs. Each pair therefore receives equal weight regardless of its frequency. For precision, we calculated the matching proportion for each distinct asserted field–value pair, averaged these proportions within each field, then averaged the field scores across fields for which both methods asserted at least one value.

**Statistical comparison.** All comparisons are paired, with records included only when both methods produce an output, ensuring that the methods are evaluated on identical inputs. We report 95% confidence intervals for both units of analysis but perform hypothesis testing only on deduplicated scores because repeated field–value pairs make field-level observations non-independent. Field-level intervals are estimated using a cluster bootstrap over records, and deduplicated intervals and *p*-values are obtained using bootstrap and paired permutation tests, respectively, with 10,000 resamples.

For deduplicated comparisons, the pairing unit differs between recall and precision. Recall is paired by distinct reference-standard field–value pairs because both methods share the same reference standard. Precision cannot be paired by distinct assertions because each method may produce a different set of values; we therefore pair the per-field deduplicated precision scores defined above using the template fields shared by both methods.

*2.7 Experiment setting*

We built the agent workflow using the LangChain (v1.2) and LangGraph (v1.0) libraries, which provide a framework for chaining LLM calls with tool invocations in a structured, repeatable manner [23,24]. We used the Langfuse platform for debugging, allowing us to inspect the full sequence of messages exchanged between the LLM and its tools, as well as token consumption and execution time [25].

We used GPT-5.6-Terra for both ARMS and the prompt-only baseline because it supports both reasoning and tool use. The model was run with high reasoning effort without setting a seed or maximum output length. Each method was run separately on the records from each of the 12 assay types, yielding 24 runs (2 methods × 12 assay types), with each of the 839 legacy records processed independently. After each run was completed, we stored the resulting metadata files locally and applied the evaluation metrics and statistical tests described above.

## 3. Results

*3.1 Field-level scoring*

Pooled across all records and fields, ARMS substantially improved both precision and recall compared with prompt engineering. Precision increased from 0.56 [0.56, 0.57] for the baseline approach to 0.93 [0.93, 0.94] for ARMS, while recall increased from 0.51 [0.50, 0.51] to 0.85 [0.84, 0.85]. As anticipated by the field categories defined in Section 2.6, the magnitude of the improvement differed according to whether values had to be resolved against an ontology.

The largest gains occurred on ontology-constrained fields, where the two methods differ most directly in how values are resolved. As described in Sections 2.2 and 2.3, ARMS queries BioPortal at runtime to find valid ontology terms, whereas the baseline must resolve the same values using the LLM's pretrained knowledge. On these fields, precision increased from 0.36 [0.35, 0.38] to 0.92 [0.91, 0.93] and recall from 0.33 [0.32, 0.34] to 0.80 [0.79, 0.81]. By comparison, on non-ontology-constrained fields, precision increased from 0.70 [0.70, 0.71] to 0.94 [0.94, 0.95] and recall from 0.63 [0.62, 0.63] to 0.88 [0.87, 0.88].

The pooled pattern was also consistent across assay types. ARMS scored higher than the baseline method on both precision and recall in both field categories shown in Figure 3, as well as overall, for all 12 assay types. The strongest separation again occurred on ontology-constrained fields. Across assays, the precision of the prompt-engineering method ranged from 0.09 to 0.56 and recall from 0.09 to 0.53, whereas the ARMS precision ranged from 0.82 to 1.00 and recall from 0.62 to 1.00. The largest gains occurred in assays where the baseline method performed most poorly. For Auto-fluorescence, precision increased from 0.09 [0.06, 0.12] to 1.00 [1.00, 1.00] and recall from 0.09 [0.06, 0.11] to 0.96 [0.94, 0.97]. For Histology, precision increased from 0.20 [0.16, 0.24] and recall from 0.21 [0.16, 0.25], both to 1.00 [1.00, 1.00]. For Lightsheet, both measures increased from 0.25 [0.17, 0.33] to 1.00 [1.00, 1.00]. On Histology, Lightsheet, and MIBI, ARMS reproduced every ontology-constrained reference-standard value exactly.

ARMS also improved performance on non-ontology-constrained fields, although the gains were smaller than those observed on ontology-constrained fields. These fields do not require ontology lookup, but ARMS still has access to the complete CEDAR template specification, whereas the baseline receives only the target field names and the ontology names provided in its prompt. The baseline therefore started from a higher level of performance, with precision ranging from 0.52 to 0.86 and recall from 0.40 to 0.80 across assays. ARMS reached precision of 0.84 to 1.00 and recall of 0.75 to 0.99. Correspondingly, the narrowest margins occurred where the baseline method was already strongest. For metadata related to multiplexed ion-beam imaging (MIBI), for example, precision increased from 0.86 [0.86, 0.86] to 0.93 [0.92, 0.93] and recall from 0.80 [0.80, 0.80] to 0.91 [0.91, 0.92]. The smaller but consistent gains in this category suggest that access to the full template specification also contributes to better metadata rectification.

*3.2 Deduplicated scoring*

For ontology-constrained fields, the advantage remained large under deduplicated scoring. Across 200 distinct reference-standard field–value pairs, recall increased by 0.377 [0.302, 0.449], from 0.297 for the baseline method to 0.674 for ARMS ($p = 0.0001$). Across 101 template fields, precision increased by 0.556 [0.485, 0.627], from 0.322 to 0.878 ($p = 0.0001$). These gains show that ARMS was not simply more consistent at resolving frequently recurring ontology values but also made more accurate assertions across the diversity of ontology values represented in the corpus than the baseline method.

The non-ontology-constrained fields showed a different pattern. Deduplicated precision remained higher for ARMS, increasing by 0.179 [0.116, 0.247] across 141 fields, from 0.681 to 0.860 ($p = 0.0001$). Deduplicated recall, however, was already high in both cases and did not differ significantly between methods. Across 1,486 distinct reference-standard field–value pairs, recall was 0.848 for the baseline method and 0.843 for ARMS, not a significant difference ($p = 0.6641$). This contrasts with the field-level recall increase from 0.63 to 0.88 for the same field category. Taken together, the two analyses show that ARMS correctly reproduces more non-ontology-constrained field instances across the corpus, but without a corresponding improvement in deduplicated recall. Its field-level recall advantage in this category therefore comes primarily from consistent correction of values that recur across records.

*3.3 Ablation and robustness analyses*

To distinguish the contributions of ARMS's two external knowledge sources, we performed ablation experiments that provided either the complete CEDAR template without terminology search or BioPortal terminology search without the complete template. Template retrieval accounted for most of the improvement on non-ontology-constrained fields, increasing precision and recall from 0.704 and 0.626 for the baseline to 0.944 and 0.881, respectively, closely matching full ARMS (0.942 and 0.878). In contrast, terminology search produced its largest gains on ontology-constrained fields, increasing precision from 0.365 to 0.803 and recall from 0.331 to 0.572. Neither capability alone matched full ARMS on ontology-constrained fields, for which precision and recall reached 0.919 and 0.799 when both were available. These results indicate complementary contributions: the complete template primarily supports field interpretation and constraint-aware transformation, whereas terminology search supports resolution of valid ontology values. Detailed assay-level results are provided in Supplementary Section B.

Because LLM outputs are nondeterministic, we also assessed whether the observed performance was stable across repeated executions. We repeated the baseline and ARMS three times using GPT-5.6-Terra and observed little variation in precision and recall, with ARMS maintaining its performance advantage across runs. Most individual field predictions were also identical across the three runs, with particularly high consistency for ARMS on ontology-constrained fields (Supplementary Section C).

We further examined whether the observed performance depended on the choice of underlying model by applying the ARMS workflow using three GPT-5.6 model tiers and a locally hosted open-weight model on four representative assays. Performance on ontology-constrained fields was broadly similar across models, whereas greater variation occurred on non-ontology-constrained fields. These exploratory single-run comparisons suggest that ARMS can operate with different underlying models, although broader repeated experiments would be needed to establish differences among models (Supplementary Section D).

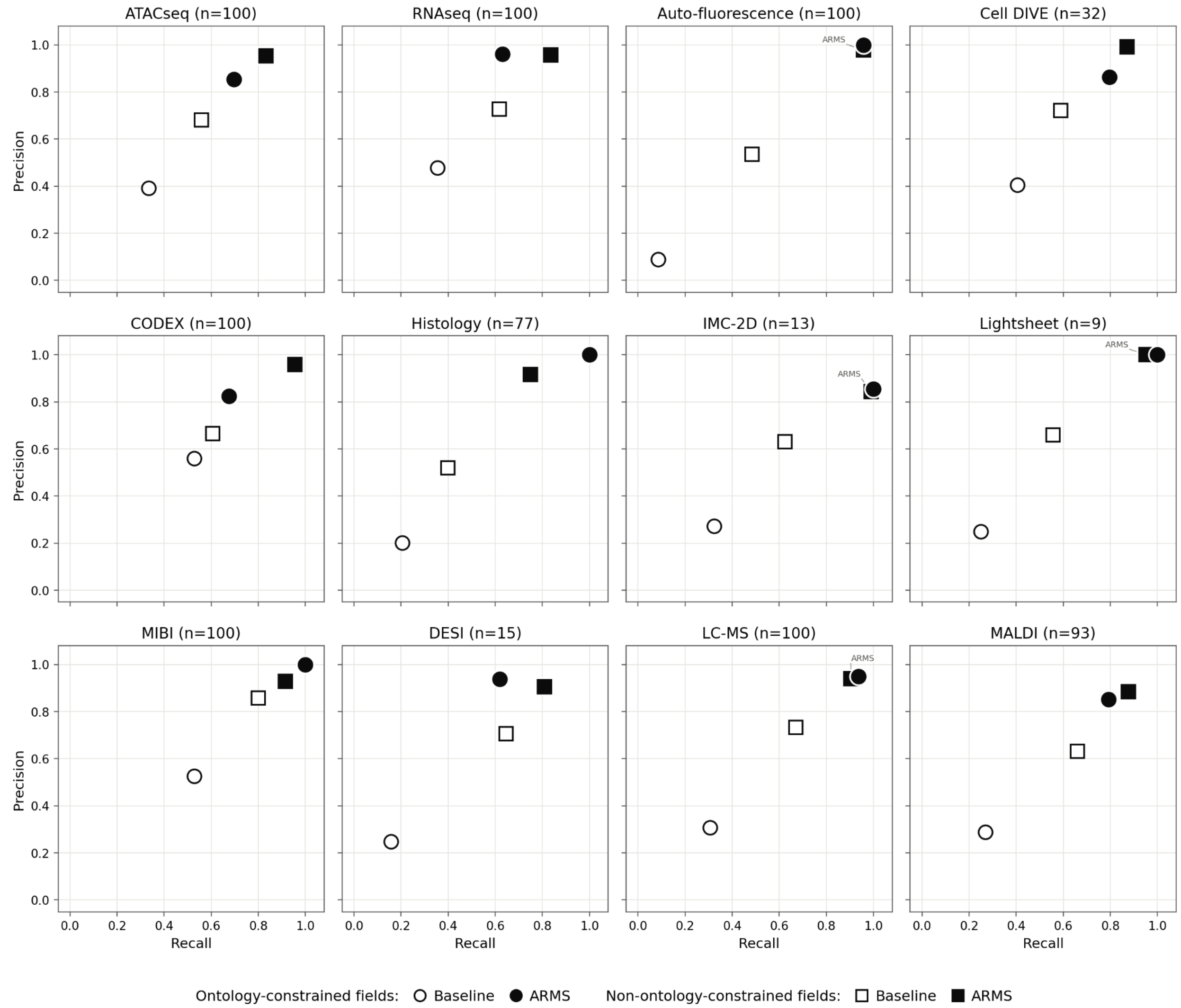

**Figure 3. Precision and recall by assay type and field category**. Each panel is one assay type, with the number of records in parentheses; recall is on the horizontal axis and precision on the vertical axis. Open markers are the prompt-only LLM baseline and filled markers the ARMS agent, and marker shape denotes the metadata field category. Points nearer the upper right indicate better performance. In every panel the filled markers sit above and to the right of the open ones, indicating ARMS achieved higher precision and recall for every assay, with the largest shift occurring for ontology-constrained fields, indicating that these fields benefited most from ARMS.

### *3.4 Error analysis*

We use "disagreement" to denote a field-level mismatch between a method's output and the reference standard that takes the form of an omission, substitution, or insertion. We examined disagreements for both methods across the same 839 records, counting repeated disagreements once within each assay by field, reference value, asserted value, and error subcategory. The baseline produced 715 distinct disagreements: 491 substitutions (69%), 138 omissions (19%), and 86 insertions (12%), while ARMS produced 461: 98 substitutions (21%), 304 omissions (66%), and 59 insertions (13%). Thus, baseline disagreements predominantly involved asserting a different value, whereas ARMS disagreements predominantly involved leaving a reference-valued field blank.

The predominance of omissions in ARMS is consistent with the reported precision–recall pattern, in which ARMS's precision exceeds recall. Omissions account for two thirds of the disagreements and fall into two categories based on whether the expected value is present in the legacy record or supplied by the human curator using information beyond the record. In 229 cases, the value was present in the legacy record, but ARMS did not use it and left the field empty.

In the remaining 75 cases, ARMS left the field empty because the legacy record did not establish the expected value, which may depend on information outside the record, such as in protocol documents or product catalogs. One example is in RNAseq, a legacy record identified the “10x Chromium Single Cell 3′ v3.1” kit but did not specify which sequencing read contained the UMI. The reference populated `umi_read` with “Read 1 (R1),” which requires knowledge of the kit’s read structure beyond the recorded metadata. Similarly, an ATACseq legacy record contained transposition kit information but did not identify the sample indexing kit. The reference populated the ontology-constrained `sample_indexing_kit` field with the appropriate permissible value from the curator.

Closer inspection of the 229 cases shows that most omissions occurred because ARMS deliberately rejected a legacy value that did not meet the template specification, rather than because it failed to recognize information in the legacy record. The largest example is the `parent_sample_id` field (205 cases), where the reference standard retains legacy identifiers such as “STAN0003-SI-2-12” even though the value does not match the identifier pattern defined in the specification. Similar conflicts occur when a legacy value is not sufficiently specific to select among the template’s permissible values. For example, the legacy value “NovaSeq 6000 S4 Reagent” for `sequencing_reagent_kit` did not establish the version, cycle count, or product number needed to distinguish among the label candidates returned by the term search tool. In these cases, leaving the field empty follows the system prompt: ARMS does not assert a value that violates the specification, nor does it attempt to choose among equally plausible alternatives. Some measured omissions therefore reflect deliberate adherence to the target specification rather than a failure to recognize information in the legacy record.

Substitutions account for 98 cases where ARMS supplied values that differed from the reference standard. Of these, 23 are near matches that differ in formatting or string structure, such as directory-path notation and sequencing read-format delimiters. These differences are counted as disagreements under exact matching, which is appropriate because the corresponding fields require specific string patterns. Another 46 cases are classified as potential field-mapping errors. Among these are cases in MALDI and Histology, where ARMS populated `preparation_protocol_doi` from `section_prep_protocols_io_doi`, whereas the reference used `protocols_io_doi`. The remaining 29 cases involve strings supplied by ARMS that differ in content from the reference and are not found in the legacy record.

Insertions account for 59 distinct cases in which ARMS populated a field that was blank in the reference standard. In 48 cases, ARMS copied a value from a semantically related legacy field to the target field, whereas the reference standard left the target field blank. Examples include populating `number_of_pcr_cycles_for_indexing` from `library_pcr_cycles` in ATACseq, `number_of_total_imaging_rounds` from `number_of_cycles` in CODEX, and `lab_id` from `library_id`. In the remaining 11 cases, ARMS derived a value from other information in the legacy record rather than directly copying an existing value. For example, in IMC-2D, the reference standard left `total_run_time_value` and `total_run_time_unit` blank, while ARMS calculated the duration from the available `start_datetime` and `end_datetime` values. Supplementary Section E lists the subcategories used to group each disagreement with their definitions and the analysis summary results.

Because the primary evaluation measures agreement with expert-curated reference records, it assumes that the reference values represent the target standard. However, the error analysis identified cases in which reference values themselves violated the current CEDAR template constraints, meaning that disagreement with the reference does not necessarily indicate an ARMS error. To assess whether such nonconformant reference values affected the measured performance of ARMS, we performed a secondary evaluation excluding fields with objectively verifiable violations of the CEDAR template constraints. We limited these exclusions to violations that could be determined without subjective reassessment of the curated records, such as datatypes, regular-expression patterns, and closed permissible-value lists. If any populated reference value violated one of these constraints, the field was excluded from all records of that assay for both the baseline and ARMS. After excluding 33 assay–field combinations, ARMS’s overall precision increased from 0.93 to 0.95 and recall from 0.85 to 0.89, whereas baseline performance remained similar, with precision changing from 0.56 to 0.55 and recall from 0.51 to 0.50. These results indicate that the original reference standard modestly underestimated ARMS’s performance when evaluation is restricted to fields consistent with the target template constraints. The complete exclusion procedure, audit results, and re-evaluated performance are provided in Supplementary Section F.

## 4. Discussion

Biomedical investigators increasingly hope to expose AI methods to the vast amount of data available in online repositories, yet they remain stymied by the remarkably poor quality of the metadata that would allow humans—and machines—to make sense of what past investigators have done in their scientific experiments. Although many repositories provide minimal metadata standards that they would like investigators to use when curating their datasets, there is typically very little enforcement of such standards. Thus, as expectations remain high that AI will provide enormous opportunities to re-analyze existing datasets and to provide a pathway toward new scientific discoveries, very few online datasets are “AI ready.”

The advent of LLM technology gave considerable hope that workers in informatics could point commercial technology such as GPT and Claude at online repositories and clean up legacy metadata effortlessly. Although that expectation did not pan out, our group showed that providing LLMs with knowledge of the target standards to which metadata should adhere as part of their prompts led to substantial improvement in the ability of LLMs to curate metadata records [9,10]. In the work presented in this paper, we show how an agentic AI system, ARMS, is capable of real-time queries to external resources such as the BioPortal ontology repository [19], providing an even better solution.

Our results show that ARMS substantially outperformed an LLM that relied on prompt-engineering communication of standards for rectifying legacy metadata from the HuBMAP repository. We demonstrated the largest gains when values had to be resolved against controlled terminologies. These findings indicate that LLMs achieve metadata standardization more reliably when they can consult authoritative domain resources rather than relying solely on pretrained knowledge. The remaining differences in performance also identify important boundaries for automated metadata standardization.

**First, access to authoritative knowledge substantially reduces the difficulty of standardizing legacy values to valid ontology terms.** The strongest evidence comes from the contrast between the performance of two approaches when viewed in terms of metadata field categories. In the prompt-only baseline, precision was 0.36 on ontology-constrained fields compared with 0.70 on non-ontology-constrained fields. With ARMS, this gap nearly disappeared, with precision of 0.92 and 0.94, respectively. The same pattern occurred across all 12 assay types. This result suggests that the baseline’s difficulty is not necessary a lack of semantic understanding, but a lack of reliable access to the external knowledge needed to translate that understanding into valid, constraint-compliant metadata assertions. Naming an ontology or branch in a prompt tells the model where an answer belongs, but it does not give the model reliable access to the terms that currently belong there. Retrieving and querying those resources at execution time closes that gap.

The ablation analysis helps explain why ARMS improves both field categories despite their different constraints. Access to the complete CEDAR template accounted for most of the improvement on non-ontology-constrained fields, whereas terminology search contributed primarily to ontology-constrained value resolution. Neither capability alone reproduced the full improvement on ontology-constrained fields, indicating that template-level knowledge and authoritative terminology access play complementary roles. ARMS therefore benefits not simply from adding external information to the LLM, but from providing different forms of machine-actionable knowledge at the points in the standardization process where they are needed.

**Second, the improved performance reflects either more consistent handling of recurring values or the ability to standardize a wider range of distinct values.** When comparing ARMS to the prompt-engineering approach, field-level performance increased substantially across both the ontology-constrained and non-ontology-constrained categories. Deduplication, however, reveals two different mechanisms behind these gains. First, for ontology-constrained fields, the advantage remained large after each distinct value was counted once, indicating ARMS is robust at resolving a wider range of values that must conform to controlled terminologies rather than simply correcting frequently recurring terms. Second, for non-ontology-constrained fields, deduplicated precision increased but recall was essentially unchanged, indicating ARMS makes more accurate assertions but remains limited in resolving distinct values for fields that allow a wider range of possible values.

The additional experiments also provide evidence about the practical robustness of the ARMS workflow. Across three repeated executions with GPT-5.6-Terra, precision and recall varied little, and ARMS maintained its performance advantage despite the nondeterministic behavior of the underlying LLM. ARMS also produced more consistent field-level answers across runs, particularly for ontology-constrained fields, suggesting that grounding term selection in template constraints and terminology searches can make standardization more repeatable. In a separate exploratory comparison, ARMS operated with three GPT-5.6 model tiers and a locally hosted open-weight model, with broadly similar performance on ontology-constrained fields across the four assays tested. Although broader repeated comparisons are needed before drawing conclusions about differences among models, these results suggest that the benefits of the ARMS workflow are not necessarily tied to a particular LLM.

Approaches such as ARMS depend on the availability of appropriate standards for descriptive metadata. Although FAIRsharing [26] documents many biomedical metadata reporting guidelines, further work is needed to define standards that describe experimental conditions, samples, and methods across scientific domains. CEDAR represents these standards as machine-actionable templates that support prospective dataset annotation [6]. Our results show that the same specifications can guide an agent in retrospectively standardizing legacy metadata more effectively than the prompt-engineering approach evaluated here.

The same agentic design can be extended beyond template specifications and ontologies to incorporate other external resources needed for metadata standardization. CEDAR already allows developers to associate the metadata field values with naming authorities such as ORCID, RRID, and ROR. Although none of the HuBMAP metadata standards for the datasets on which we tested our approach references such external resources, it is clear that an agentic system would benefit from the ability to query them in real time when indicated. Such extensions reinforce a broader advantage of the agentic approach: Metadata standards and controlled terminologies evolve independently of the models that use them. Encoding increasingly detailed representations of these resources into prompts therefore offers only a temporary solution. By retrieving the current specification at execution time, the model can instead focus its reasoning on interpreting heterogeneous legacy metadata in context.

### *4.1 Limitations*

This study has several limitations. First, ARMS can standardize only information that can be recovered from the legacy record or from the external resources available through its current tool set. Some reference-standard values depended on experimental protocols, product documentation, or other information unavailable to the agent. Extending ARMS with access to additional authoritative resources could address some of these cases. Second, the expert-curated records provide a practical reference standard but are not an infallible ground truth. Our error analysis identified cases in which historical curation decisions conflicted with current template constraints, meaning that some measured disagreements do not necessarily represent errors by ARMS. To assess the effect of these inconsistencies, we re-evaluated our methods after excluding fields with non-compliant reference values. Third, our evaluation was conducted on HuBMAP metadata across 12 assay types using CEDAR-based specifications. Although these assays represent a range of metadata structures and constraints, the evaluation was limited to a single biomedical data repository and metadata framework. Further evaluation on other repositories, metadata standards, and scientific domains is needed to determine how well the approach generalizes beyond this setting.

## 5. Conclusion

Non-standard descriptive metadata remain a serious barrier to making research datasets findable, accessible, interoperable, and reusable, and addressing the problem at scale requires more than manual curation. Previous work showed that metadata specifications can guide an LLM toward better standardization when provided in a prompt. Our results extend that finding by showing that LLMs can standardize legacy metadata more effectively when they can retrieve complete metadata specifications and query authoritative terminology resources during execution. Our findings show that machine-actionable metadata standards can serve not only as specifications for human curation but also as executable knowledge for AI-assisted standardization. Separating authoritative domain knowledge from model reasoning provides a scalable and maintainable foundation for metadata standardization as scientific standards continue to evolve.

## Funding

This work was supported, in part, by the U.S. National Library of Medicine under grant R01 LM013498; by the U.S. National Institutes of Health Common Fund under award OT2 OD033759; and by the U.S. National Institute of General Medical Sciences under grant U24 GM143402. The funders had no role in study design, data collection and analysis, the writing of the report, or the decision to submit the article for publication.

## CRediT authorship contribution statement

Josef Hardi: Conceptualization, Methodology, Software, Formal analysis, Investigation, Visualization, Writing - original draft. Martin J. O'Connor: Conceptualization, Validation, Supervision, Writing - review and editing. Marcos Martínez-Romero: Conceptualization, Supervision, Writing - review and editing. Jean G. Rosario: Data curation, Resources, Validation. Stephen A. Fisher: Data curation, Resources, Validation. Mark A. Musen: Conceptualization, Funding acquisition, Supervision, Writing - review and editing.

## Declaration of competing interest

The authors declare that they have no known competing financial interests or personal relationships that could have appeared to influence the work reported in this paper.

## Data and code availability

- The experimental data and source code for ARMS and the evaluation methods are available on GitHub at https://github.com/musen-lab/metadata-standardization-agent.
- A Jupyter experiment notebook is provided to rerun the experiments and reproduce all reported data analyses.
- Model settings, the system and user prompts used in the experiments are documented in the experiment note.

## Declaration of generative AI and AI-assisted technologies in the manuscript preparation process

During the preparation of this work, the authors used Claude Code in order to produce the data analysis tools and the Jupyter notebook. After using this tool, the authors reviewed and edited the source code as needed and took full responsibility for the content of the published article.

## References

[1] M.D. Wilkinson, M. Dumontier, I.J.J. Aalbersberg, G. Appleton, M. Axton, A. Baak, N. Blomberg, J.-W. Boiten, L.B. da Silva Santos, P.E. Bourne, J. Bouwman, A.J. Brookes, T. Clark, M. Crosas, I. Dillo, O. Dumon, S. Edmunds, C.T. Evelo, R. Finkers, A. Gonzalez-Beltran, A.J.G. Gray, P. Groth, C. Goble, J.S. Grethe, J. Heringa, P.A.C. ’t Hoen, R. Hooft, T. Kuhn, R. Kok, J. Kok, S.J. Lusher, M.E. Martone, A. Mons, A.L. Packer, B. Persson, P. Rocca-Serra, M. Roos, R. van Schaik, S.-A. Sansone, E. Schultes, T. Sengstag, T. Slater, G. Strawn, M.A. Swertz, M. Thompson, J. van der Lei, E. van Mulligen, J. Velterop, A. Waagmeester, P. Wittenburg, K. Wolstencroft, J. Zhao, B. Mons, The FAIR Guiding Principles for scientific data management and stewardship, Sci. Data 3 (2016) 160018. https://doi.org/10.1038/sdata.2016.18.

[2] R.S. Gonçalves, M.A. Musen, The variable quality of metadata about biological samples used in biomedical experiments, Sci. Data 6 (2019) 190021. https://doi.org/10.1038/sdata.2019.21.

[3] A. Brazma, P. Hingamp, J. Quackenbush, G. Sherlock, P. Spellman, C. Stoeckert, J. Aach, W. Ansorge, C.A. Ball, H.C. Causton, T. Gaasterland, P. Glenisson, F.C.P. Holstege, I.F. Kim, V. Markowitz, J.C. Matese, H. Parkinson, A. Robinson, U. Sarkans, S. Schulze-Kremer, J. Stewart, R. Taylor, J. Vilo, M. Vingron, Minimum information about a microarray experiment (MIAME)—toward standards for microarray data, Nat. Genet. 29 (2001) 365–371. https://doi.org/10.1038/ng1201-365.

[4] M.A. Musen, M.J. O’Connor, E. Schultes, M. Martínez-Romero, J. Hardi, J. Graybeal, Modeling community standards for metadata as templates makes data FAIR, Sci. Data 9 (2022) 696. https://doi.org/10.1038/s41597-022-01815-3.

[5] R.S. Gonçalves, M.J. O’Connor, M. Martínez-Romero, A.L. Egyedi, D. Willrett, J. Graybeal, M.A. Musen, The CEDAR Workbench: An ontology-assisted environment for authoring metadata that describe scientific experiments, in: The Semantic Web – ISWC 2017, Springer, 2017: pp. 103–110. https://doi.org/10.1007/978-3-319-68204-4_10.

[6] M.A. Musen, M.J. O’Connor, J. Hardi, M. Martínez-Romero, Knowledge engineering for open science: Building and deploying knowledge bases for metadata standards, AI Mag. 47 (2026). https://doi.org/10.1002/aaai.70048.

[7] M.A. Musen, C.A. Bean, K.-H. Cheung, M. Dumontier, K.A. Durante, O. Gevaert, A. Gonzalez-Beltran, P. Khatri, S.H. Kleinstein, M.J. O’Connor, Y. Pouliot, P. Rocca-Serra, S.-A. Sansone, J.A. Wiser, CEDAR team, The center for expanded data annotation and retrieval, J. Am. Med. Inform. Assoc. 22 (2015) 1148–1152. https://doi.org/10.1093/jamia/ocv048.

[8] K. Long, K. Gravel-Pucillo, L. Waldron, S. Davis, S. Oh, Large-scale manual curation and harmonization of metadata from metagenomic and cancer genomic repositories: challenges and solutions, Database (Oxford) 2026 (2026) baag027. https://doi.org/10.1093/database/baag027.

[9] S.S. Sundaram, B. Solomon, A. Khatri, A. Laumas, P. Khatri, M.A. Musen, Structured knowledge base enhances effective use of large language models for metadata curation, AMIA Annu. Symp. Proc. 2024 (2024) 1050–1058. https://pmc.ncbi.nlm.nih.gov/articles/PMC12099408/.

[10] S.S. Sundaram, R.S. Gonçalves, M.A. Musen, Toward total recall: Enhancing data FAIRness through AI-driven metadata standardization, Gigascience 15 (2026) giag019. https://doi.org/10.1093/gigascience/giag019.

[11] D.B. Hier, S.K. Platt, T. Obafemi-Ajayi, Predicting failures of LLMs to link biomedical ontology terms to identifiers: Evidence across models and ontologies, in: 2025 IEEE EMBS International Conference on Biomedical and Health Informatics (BHI), IEEE, 2025: pp. 1–7. https://doi.org/10.1109/bhi67747.2025.11269551.

[12] N.J. Dobbins, Generalizable and scalable multistage biomedical concept normalization leveraging large language models, Res. Synth. Methods 16 (2025) 479–490. https://doi.org/10.1017/rsm.2025.9.

[13] J.H. Caufield, H. Hegde, V. Emonet, N.L. Harris, M.P. Joachimiak, N. Matentzoglu, H. Kim, S. Moxon, J.T. Reese, M.A. Haendel, P.N. Robinson, C.J. Mungall, Structured Prompt Interrogation and Recursive Extraction of Semantics (SPIRES): a method for populating knowledge bases using zero-shot learning, Bioinformatics 40 (2024). https://doi.org/10.1093/bioinformatics/btae104.

[14] P. Lewis, E. Perez, A. Piktus, F. Petroni, V. Karpukhin, N. Goyal, H. Küttler, M. Lewis, W.-T. Yih, T. Rocktäschel, S. Riedel, D. Kiela, Retrieval-augmented generation for knowledge-intensive NLP tasks, in: Proceedings of the 34th International Conference on Neural Information Processing Systems, Curran Associates Inc., Red Hook, NY, USA, 2020: pp. 9459–9474.

[15] T. Schick, J. Dwivedi-Yu, R. Dessì, R. Raileanu, M. Lomeli, E. Hambro, L. Zettlemoyer, N. Cancedda, T. Scialom, Toolformer: Language models can teach themselves to use tools, Advances in Neural Information Processing Systems 36 (2023) 68539–68551. https://doi.org/10.52202/075280-2997.
[16] S. Yao, J. Zhao, D. Yu, N. Du, I. Shafran, K.R. Narasimhan, Y. Cao, ReAct: Synergizing Reasoning and Acting in Language Models, in: International Conference on Learning Representations, 2023. https://mlanthology.org/iclr/2023/yao2023iclr-react/.
[17] S.G. Patil, T. Zhang, X. Wang, J.E. Gonzalez, Gorilla: Large Language Model Connected with Massive APIs, in: A. Globerson, L. Mackey, D. Belgrave, A. Fan, U. Paquet, J. Tomczak, C. Zhang (Eds.), Advances in Neural Information Processing Systems, Curran Associates, Inc., 2024: pp. 126544–126565. https://doi.org/10.52202/079017-4020.
[18] S. Chen, T. Alrashed, A. Halevy, N. Noy, Do data agents need semantic metadata? A comparative study in agentic data retrieval, arXiv [Cs.IR] (2026). https://doi.org/10.48550/arXiv.2605.28787.
[19] J. Vendetti, N.L. Harris, M.V. Dorf, A. Skrenchuk, J.H. Caufield, R.S. Gonçalves, J.B. Graybeal, H. Hegde, T. Redmond, C.J. Mungall, M.A. Musen, BioPortal: an open community resource for sharing, searching, and utilizing biomedical ontologies, Nucleic Acids Res. 53 (2025) W84–W94. https://doi.org/10.1093/nar/gkaf402.
[20] HuBMAP Consortium, The human body at cellular resolution: the NIH Human Biomolecular Atlas Program, Nature 574 (2019) 187–192. https://doi.org/10.1038/s41586-019-1629-x.
[21] S.A. Fisher, J. Hardi, R. Morgan, E. Nordgren, P.M. Kant, B. Honick, J. Rosario, M.J. O'Connor, P.D. Blood, DCWG Members, J.C. Silverstein, M.A. Musen, The HuBMAP framework for advancing data FAIRness, bioRxiv (2026) 2026.06.01.728946. https://doi.org/10.64898/2026.06.01.728946.
[22] Anthropic, Introducing the Model Context Protocol, (2024). https://www.anthropic.com/news/model-context-protocol (accessed August 28, 2026).
[23] LangGraph, (2024). https://github.com/langchain-ai/langgraph (accessed August 28, 2026).
[24] H. Chase, LangChain, (2022). https://github.com/langchain-ai/langchain (accessed August 28, 2026).
[25] Langfuse, (2023). https://langfuse.com/ (accessed August 28, 2026).
[26] S.-A. Sansone, P. McQuilton, P. Rocca-Serra, A. Gonzalez-Beltran, M. Izzo, A.L. Lister, M. Thurston, FAIRsharing Community, FAIRsharing as a community approach to standards, repositories and policies, Nat. Biotechnol. 37 (2019) 358–367. https://doi.org/10.1038/s41587-019-0080-8.

## Supplementary Materials

### *A. Reference standard dataset*

**Table A.1. Reference Dataset Breakdown.** The reference dataset contains 839 legacy metadata records paired with expert-curated versions across 12 assay types. Each assay has its own template, with 23 to 48 fields. Across the 12 templates, 150 field definitions are ontology-constrained and 225 are not. The number of records per assay ranges from 9 to 100. The ontology-constrained fields use the HuBMAP Research Attributes Value Set (HRAVS)[1], a curated value set that collects terminologies from reference biomedical ontologies such as the OBO Foundry ontologies, NCI Thesaurus, Bio-Assay Ontology (BAO), Experimental Factor Ontology (EFO), and MeSH, as well as registries such as Research Resource Identifiers (RRIDs) and Logical Observation Identifiers Names and Codes (LOINC).

| Assay name | # of Records | # of Template fields | # of Ontology-constrained fields | # of Non-ontology-constrained fields |
|---|---|---|---|---|
| ATACseq | 100 | 47 | 21 | 26 |
| RNAseq | 100 | 48 | 20 | 28 |
| Auto-fluorescence | 100 | 24 | 8 | 16 |
| Cell DIVE | 32 | 23 | 6 | 17 |
| CODEX | 100 | 25 | 9 | 16 |
| Histology | 77 | 29 | 12 | 17 |
| IMC-2D | 13 | 24 | 8 | 16 |
| Lightsheet | 9 | 23 | 8 | 15 |
| MIBI | 100 | 27 | 8 | 19 |
| DESI | 15 | 31 | 16 | 15 |
| LC-MS | 100 | 46 | 20 | 26 |
| MALDI | 93 | 28 | 14 | 14 |
| **Total** | **839** | **375** | **150** | **225** |

[1] HuBMAP Research Attributes Value Set in BioPortal: https://bioportal.bioontology.org/ontologies/HRAVS. Accessed on September 28, 2026.

*B. Ablation-run analysis*

We evaluated the contribution of ARMS's two tool groups by removing one group at a time. The *template-only* experiment retained access to the complete CEDAR metadata specification but excluded BioPortal term searches. The *search-only* experiment retained the BioPortal term-search tools but excluded retrieval of the complete template, providing field and vocabulary names as in the baseline. We ran each ablation once using GPT-5.6-Terra across 12 assays. Precision and recall were calculated for both ontology-constrained and non-ontology-constrained fields.

**Template retrieval alone.** Figure B.1 shows that access to the complete template substantially improved performance on non-ontology-constrained fields in every assay.  Pooled across all assays, precision increased from 0.704 for the baseline to 0.944, and recall increased from 0.626 to 0.881. These scores closely matched full ARMS, which achieved precision of 0.942 and recall of 0.878. The results suggest that template descriptions, enumerations, datatypes, and formatting constraints supply much of the information needed to map and normalize these values. Template retrieval also improved ontology-constrained pooled precision from 0.365 to 0.599 and recall from 0.331 to 0.554, although both remained below full ARMS at 0.919 and 0.799, respectively.

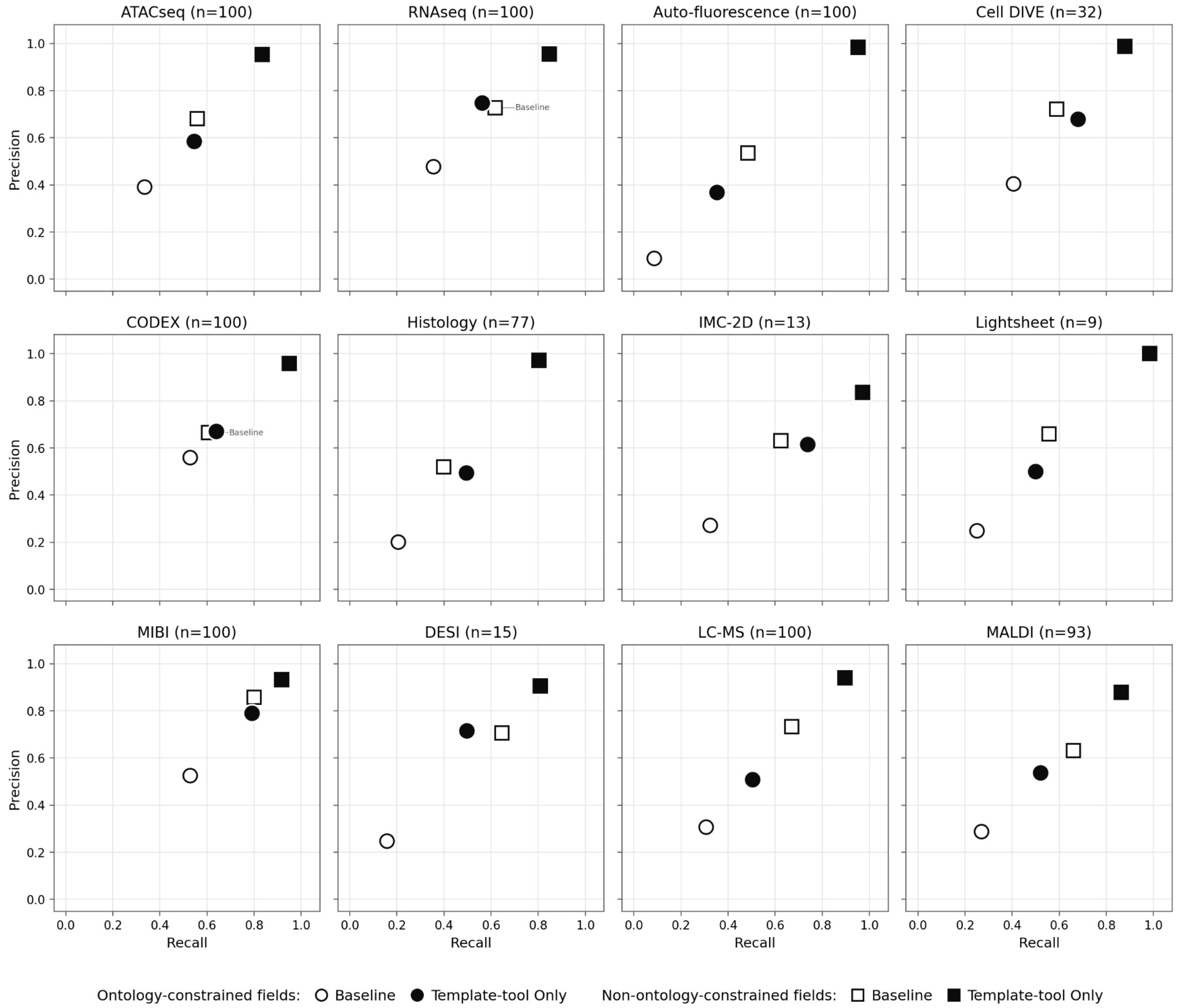

**Figure B.1. Precision and recall with template retrieval only.** Each panel shows one assay, with the number of records in parentheses. Open markers represent the baseline and filled markers the template-only condition, and marker shape denotes the metadata field category. Recall is on the horizontal axis and precision on the vertical axis. Points nearer the upper right indicate better performance.

**Ontology search alone.** Figure B.2 shows that term search primarily improved performance on ontology- constrained fields in every assay. Pooled across all assays, precision increased from 0.365 for the baseline to 0.803, while recall increased from 0.331 to 0.572. The larger improvement in precision suggests that access to ontology candidates helped the model produce more accurate term assertions. For non-ontology-constrained fields, the baseline and search-only markers closely overlap in several assays, indicating little change when ontology search was added. This pattern suggests that ontology search contributes mainly to fields requiring controlled vocabulary terms, while improvements in other fields depend more on access to the complete template.

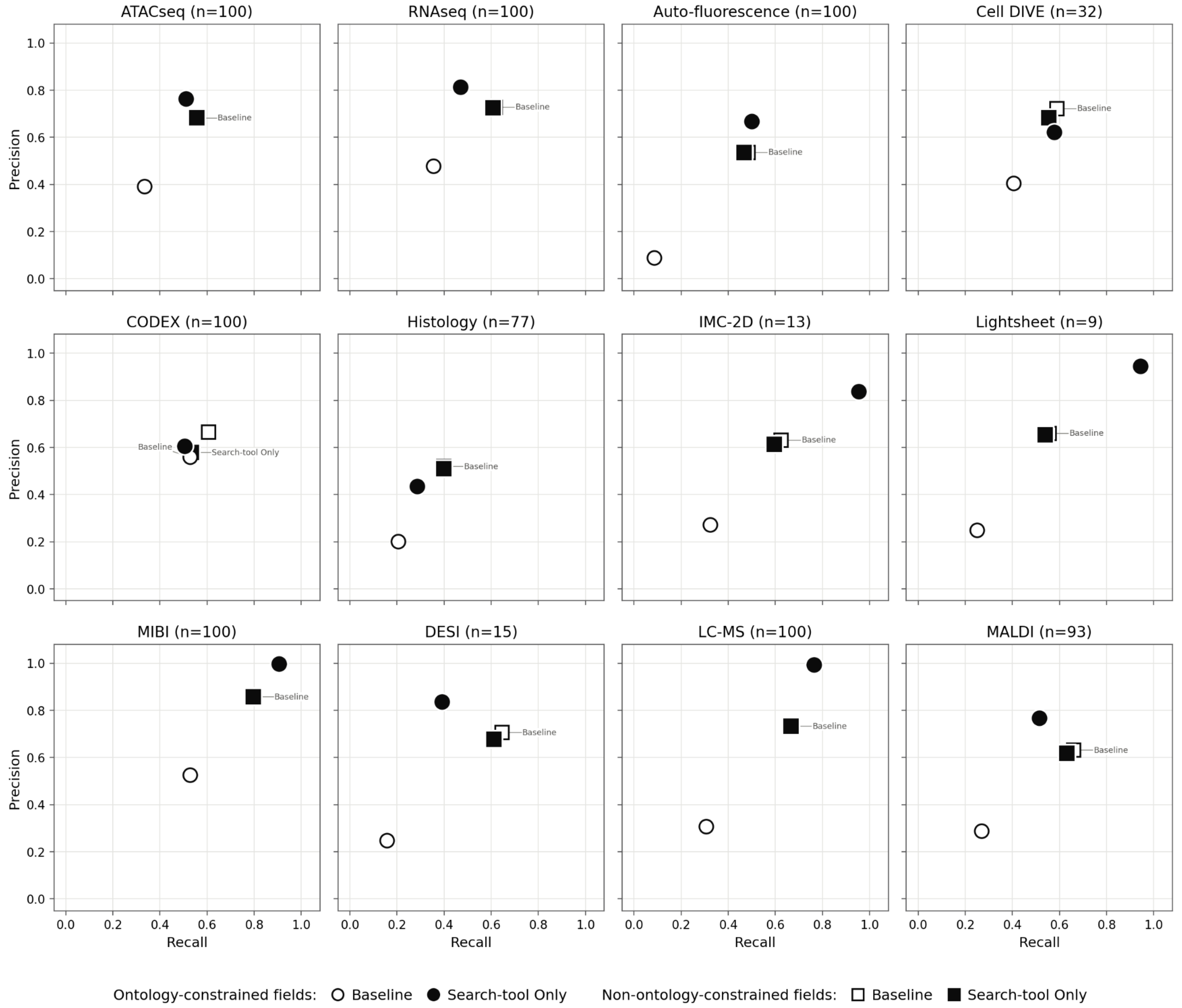

**Figure B.2. Precision and recall with ontology search only.** Each panel shows one assay, with the number of records in parentheses. Open markers represent the baseline and filled markers the search-only condition. Circles denote ontology-constrained fields, and squares denote non-ontology-constrained fields. Recall is on the horizontal axis and precision on the vertical axis; points nearer the upper right indicate better performance.

Together, these ablations suggest complementary roles for template retrieval and ontology search tools. Template retrieval accounts for most of the observed improvement on non-ontology-constrained fields, while ontology search provides a larger improvement in the precision of ontology-constrained values. Combining both tool groups yielded the highest pooled ontology-constrained precision and recall, suggesting that the complete specification helps guide field interpretation while terminology searches support term resolution. These findings help explain the performance of the full ARMS workflow examined in the repeated-run analysis. Because each ablation was run only once, small differences between the ablation runs and full ARMS performance should therefore be treated as tentative.

*C. Repeated-run analysis*

We ran the baseline and ARMS three times on the same GPT-5.6-Terra model and assay records, and assessed repeatability in two ways. We compared precision and recall across the baseline's and ARMS's runs to see how much their scores changed. We also measured how often all three runs gave the same answer for a field in a record.

**Variation in precision and recall.** Figure C.1 shows that both baseline and ARMS produced similar precision and recall across their three runs, with the markers generally forming tight clusters for each field type. For each assay and metric, we calculated the range across the three runs, then averaged these ranges across assays and precision and recall. The mean range was 3.24 percentage points for the baseline and 0.61 for ARMS on ontology-constrained fields, and 0.70 and 1.24 points, respectively, on non-ontology-constrained fields. Thus, run-to-run changes in the performance were generally small for both methods, while ARMS maintained higher precision and recall. These stable scores do not necessarily mean that every field received the same answer across runs; the next analysis examines that directly.

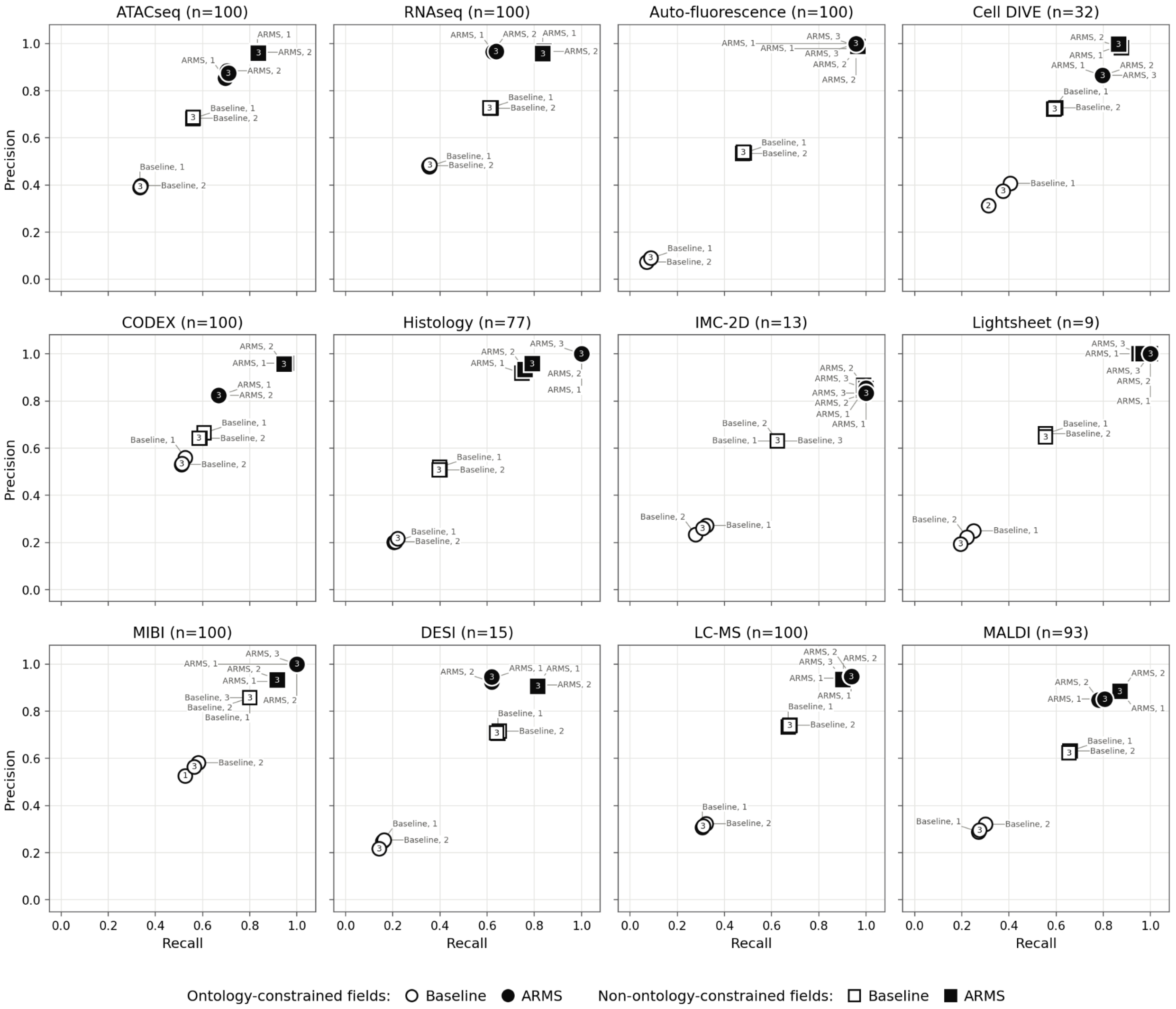

**Figure C.1. Precision and recall across three repeated runs.** Each panel is one assay type, with the number of records in parentheses; recall is on the horizontal axis and precision on the vertical axis. Open markers are the baseline and filled markers the ARMS agent, and marker shape denotes the field category. Numbers identify the runs 1 to 3. Points nearer the upper right indicate better performance. A tight clustering among the same markers indicates small variability.

**Consistency of individual answers.** For every field of a record predicted in all three runs, we compared the three outputs whether they were identical. Figure C.2 shows agreement by assay, with ontology- and non-ontology-constrained fields combined. ARMS had higher agreement in 11 of the 12 assays. Pooling all 28,257 field instances across 839 records, ARMS returned the same answer in every run for 93.5% of instances, compared with 88.4% for the baseline.

A separate analysis by field type (not shown in the figure) found that the improvement was concentrated in ontology-constrained fields. Agreement increased from 78.7% for the baseline to 95.4% for ARMS. Access to the template's vocabulary constraint and BioPortal terminology searches could make term selection more repeatable. For non-ontology-constrained fields, agreement was slightly lower for ARMS: 92.2%, compared with 94.9% for the baseline. The baseline often retains mapped legacy values as written, whereas ARMS also interprets template descriptions and formatting rules, and these additional decisions may contribute to variation across runs.

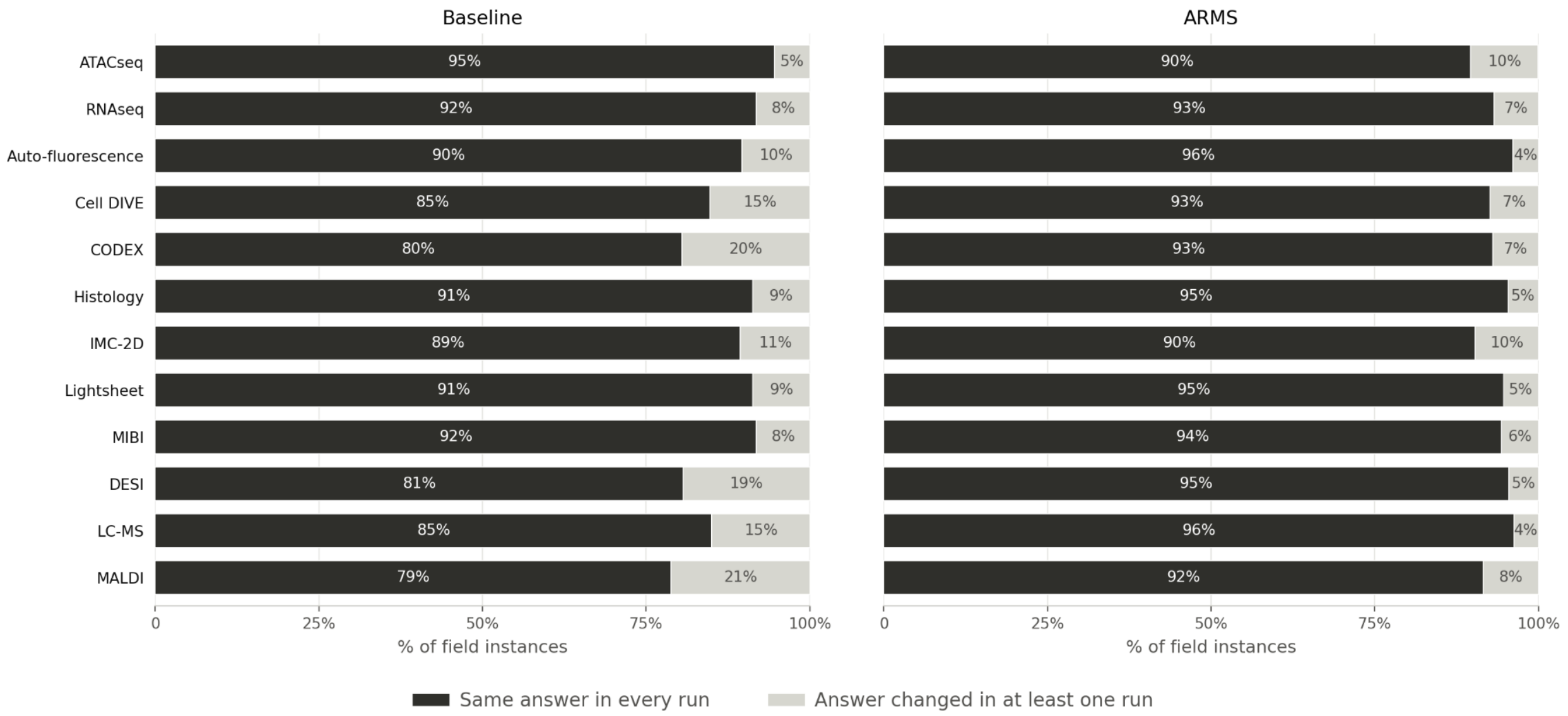

**Figure C.2. Field-level answer consistency across three repeated runs.** The vertical axis lists the assays, and the horizontal axis shows the percentage of field instances. The left and right panels show the baseline and ARMS, respectively. Each bar shows the percentage of field instances within an assay that received the same answer in all three runs (dark) or changed in at least one run (light). Agreement measures consistency independently of correctness.

Through this experiment, we found that although GPT-5.6-Terra generates responses nondeterministically, its precision and recall remained relatively stable and most field predictions were identical across the three runs. These findings suggest relatively robust behavior on this metadata standardization task under the tested configuration. The higher agreement on ontology-constrained fields with ARMS further suggests that grounding term selection in ontology search results and template vocabulary constraints may improve repeatability when using a nondeterministic LLM.

*D. Comparison with other models*

We ran the ARMS agent using four large language models: GPT-5.6-Luna, GPT-5.6-Terra, GPT-5.6-Sol, and Qwen3.8-Flash-Next. We ran each model once on CODEX (n=100), Histology (n=77), IMC-2D (n=13), and Lightsheet (n=9) records. The workflow, agent tools, input records, and reference outputs were the same across models. We calculated precision and recall separately for ontology-constrained and non-ontology-constrained fields.

We ran the GPT models with high reasoning effort without specifying a seed or maximum output length. Although the client was configured with temperature 0, inspection of the GPT-5.6 request showed that this value was not transmitted, leaving temperature at the provider default. Qwen was run with medium reasoning effort with extra thinking-mode settings[2]: `temperature`=1.0, `top_p`=0.95, `top_k`=20, `min_p`=0.0, `presence penalty`=0.0 and `repetition penalty`=1.0. The first four settings govern how the model chooses its next piece of text, and the two penalties govern repetition.

The Qwen model was served locally on an NVIDIA DGX Spark with a GB10 Grace Blackwell superchip, a 20-core ARM CPU, an integrated Blackwell GPU, and 128 GB of shared LPDDR5x memory.

**Precision and recall comparison.** Figure D.1 shows little precision and recall variation among models on ontology-constrained fields. Histology and Lightsheet show identical perfect results across all four models, while IMC-2D shows only little separation. Differences are more apparent in CODEX, where Sol recovers the most reference values at similar precision, followed by Terra, while Luna and Qwen achieve similar recall. That said, Qwen's comparable performance across these sample assays is notable because its model weights are freely available and can be run locally. In this single-run comparison, the higher GPT tiers showed no consistent advantage on ontology-constrained fields.

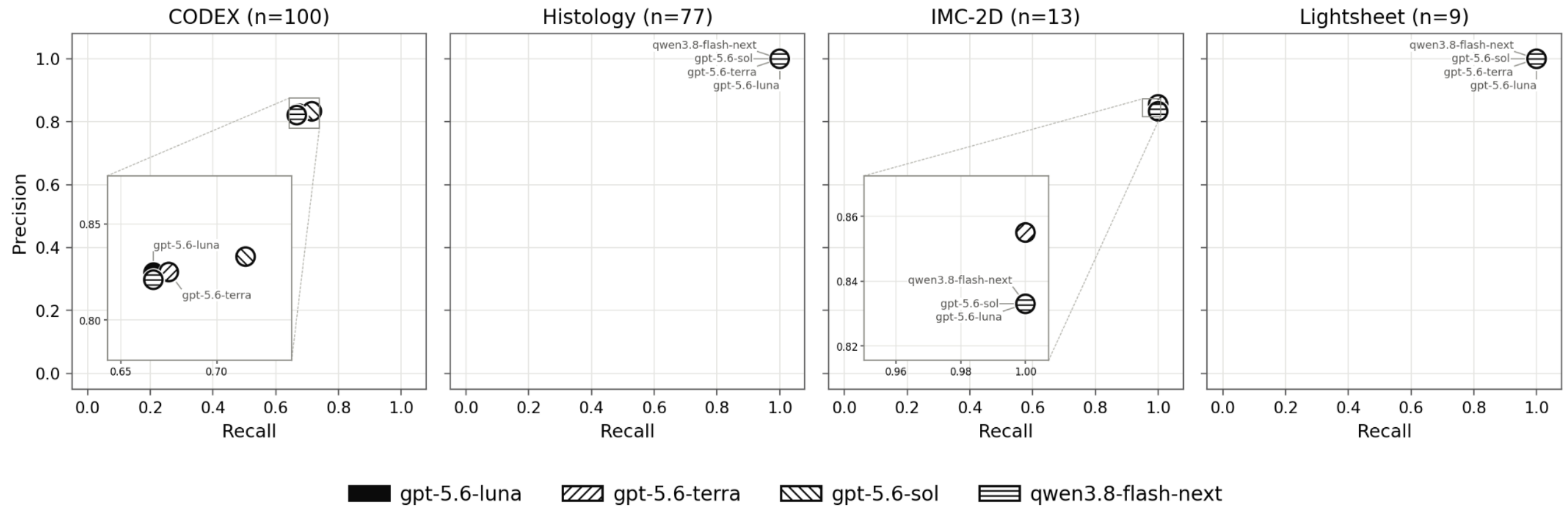

**Figure D.1. ARMS performance on ontology-constrained fields by model**. Each panel shows one assay, with a patterned marker for the ARMS run of each model. Recall is on the horizontal axis and precision on the vertical axis; points nearer the upper right indicate better performance. Field-level precision and recall were calculated for ontology-constrained fields. Insets enlarge closely spaced markers, with labels identifying overlapping markers.

By contrast, Figure D.2 shows greater variation among models on non-ontology-constrained fields. We observed that performance does not consistently follow the GPT model tiers: Sol has lower recall than Terra in every assay, while Luna, the lowest GPT tier, matches or exceeds Sol in both precision and recall across all four assays. Luna also comes close to Terra on both measures in CODEX. The local Qwen model had the highest recall in Histology. Its precision was lowest in CODEX, IMC-2D, and Lightsheet; in Histology, however, it was between Terra and Luna. The wider spread may reflect greater room for model-specific decisions when fields lack ontology guidance.

[2] Qwen3.8-Flash-Next model card on Hugging Face: https://huggingface.co/Qwen/Qwen3.8-Flash-Next. Accessed on September 28, 2026.

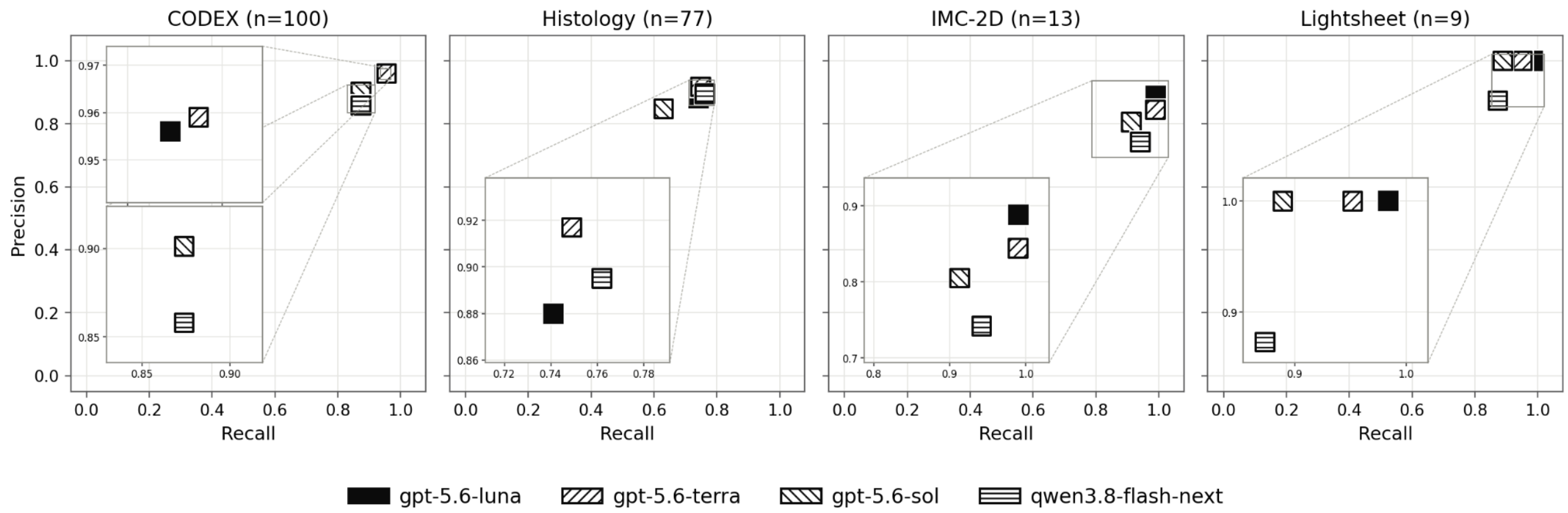

**Figure D.2. ARMS performance on non-ontology-constrained fields by model.** Each panel shows one assay, with a patterned marker for the ARMS run of each model. Recall is on the horizontal axis and precision on the vertical axis; points nearer the upper right indicate better performance. Field-level precision and recall were calculated for fields without ontology constraints. Insets enlarge closely spaced markers, with labels identifying overlapping markers.

These single-run comparisons illustrate how model choice can affect ARMS on the selected assays. Establishing a stable ranking would require repeated runs and broader assay coverage.

**Time and cost comparison.** Table D.1 shows the time spent in model calls and the recorded cost per metadata record, with latency outliers excluded from the summary. Terra had the shortest typical model-call time. Luna and Sol took somewhat longer with Sol's unfiltered CODEX traces containing several unusually long model calls. The locally hosted Qwen3.8-Flash-Next took substantially longer than the GPT models with a mean model-call time 8.16 times that of Terra.

Among models with metered API charges, Luna cost the least, followed by Terra and Sol. Qwen incurred no metered API charge, although the traces do not include the cost of operating the DGX Spark. The GPT runs used the Stanford AI API Gateway[3], which provided a 50% discount on standard OpenAI input and output rates during the experiment.

**Table D.1. Per-record model-call time and tracer-recorded cost by model**. Trace data were exported from a locally deployed Langfuse instance. For each model, rows with the same `traceId` were grouped and their `latencyMs` and `totalCost` values summed, producing 199 records before outlier exclusion. Records with outlying summed latency were excluded from the summaries using a two-sided 1.5 × interquartile range (IQR) filter. Times are in minutes and costs are in US cents. Bold text identifies Terra, the model used in the main experiment.

| Model | # of records | Time (min) *average* | Time (min) *median [min–max]* | Cost (US¢) *average* | Cost (US¢) *median [min–max]* |
|---|---|---|---|---|---|
| GPT-5.6-Luna | 194 | 1.14 | 1.14 [0.78–1.55] | 0.50 | 0.50 [0.37–0.67] |
| **GPT-5.6-Terra** | **197** | **0.81** | **0.81 [0.47–1.17]** | **7.70** | **7.65 [5.32–10.78]** |
| GPT-5.6-Sol | 180 | 1.26 | 1.24 [0.76–1.87] | 11.53 | 11.56 [7.71–14.34] |
| Qwen3.8-Flash-Next | 198 | 6.61 | 4.37 [2.10–19.92] | 0.00 | 0.00 [0.00–0.00] |

[3] Stanford AI API Gateway. https://uit.stanford.edu/service/ai-api-gateway. Accessed on September 28, 2026.

*E. Disagreement taxonomy*

We compared ARMS outputs with the curated reference standard at the field level and classified each disagreement using the taxonomy in Table E.1. We assigned a substitution when both values were populated but differed, an omission when only the reference was populated, and an insertion when only ARMS populated the output. We then assigned subcategories algorithmically using Python code. For substitutions, the algorithm first identified near matches that differ only in formatting or string structure, including differences in numeric formatting, capitalization, directory-path notation, and delimiters in sequencing read lengths. It then flagged potential wrong mappings when the asserted value was found anywhere in the legacy record. All remaining substitutions were assigned to the different-value subcategory.

For omissions, the algorithm classified cases as missed values when the reference value was present in the legacy record but ARMS left the target field blank, and as external gaps when the legacy record did not provide the reference value for the target field, suggesting that the curator may have obtained it from external resources. For insertions, cases were classified as unexpected copies when ARMS filled a field left blank in the reference with a value present in the legacy record. Unexpected fills occurred when ARMS filled such a field with a value inferred or calculated from legacy information, and it is not explicitly recorded in the legacy record.

**Table E.1. Disagreement categories and subcategories.** The three error categories—substitutions, omissions, and insertions—and definitions of their seven subcategories used to classify disagreements between ARMS outputs and the curated reference standard.

| Category | Subcategory | Definition |
|---|---|---|
| Substitution | Near match | ARMS supplies a value that differs from the reference in formatting or string structure. |
| | Wrong mapping | ARMS supplies a value that differs from the reference and is found in the legacy record, indicating a potential mapping error. |
| | Different value | ARMS supplies a value that differs from the reference and is not found in the legacy record. |
| Omission | Missed value | ARMS leaves the field blank although the expected value is available in the legacy record. |
| | External gap | ARMS leaves the field blank when the legacy record does not establish the expected value; it may depend on information from external resources. |
| Insertion | Unexpected copy | ARMS fills a field left blank in the reference with a value found in the legacy record. |
| | Unexpected fill | ARMS fills a field left blank in the reference with a value not explicitly found in the legacy record; the value may be inferred or calculated from the legacy record. |

After assigning subcategories, we grouped disagreements by subcategory, assay, target field, reference value, and ARMS value. Repeated occurrences of the same combination across records contributed one distinct case. Table E.2 shows the distinct disagreement counts by assay and subcategory. Most distinct disagreements involved non-ontology-constrained fields (358 cases; 77.7%), particularly missed values. Ontology-constrained fields accounted for 103 cases (22.3%), predominantly external gaps where the legacy records did not provide the expected values and the curator may have obtained them from external resources (Table E.3).

**Table E.2. Counts of distinct disagreements by assay and subcategory.** Counts are shown for each assay and subcategory, grouped under substitutions, omissions, and insertions. Repeated occurrences of the same disagreement across records are counted once.

| Assay name | Substitution | | | Omission | | Insertion | |
|---|---|---|---|---|---|---|---|
| | Near match | Wrong mapping | Different value | Missed value | External gap | Unexpected copy | Unexpected fill |
| ATACseq | 4 | 8 | 8 | 39 | 25 | 13 | 2 |
| RNAseq | 8 | 7 | 4 | 70 | 31 | 25 | 5 |
| Auto-fluores. | 2 | 0 | 0 | 13 | 1 | 0 | 0 |
| Cell DIVE | 0 | 3 | 0 | 12 | 3 | 0 | 0 |
| CODEX | 4 | 2 | 2 | 12 | 2 | 5 | 0 |
| Histology | 0 | 5 | 0 | 18 | 1 | 0 | 0 |
| IMC-2D | 0 | 0 | 0 | 1 | 0 | 3 | 3 |
| Lightsheet | 0 | 0 | 0 | 3 | 0 | 0 | 0 |
| MIBI | 2 | 2 | 0 | 22 | 0 | 0 | 0 |
| DESI | 0 | 3 | 2 | 6 | 5 | 0 | 0 |
| LC-MS | 2 | 9 | 10 | 22 | 5 | 1 | 1 |
| MALDI | 1 | 7 | 3 | 11 | 2 | 1 | 0 |
| **Total** | 23 | 46 | 29 | 229 | 75 | 48 | 11 |

**Table E.3. Counts of distinct disagreements by field type and subcategory.** Counts are shown for each field type and subcategory, grouped under substitutions, omissions, and insertions. Repeated occurrences of the same disagreement across records are counted once.

| Field type | Substitution | | | Omission | | Insertion | |
|---|---|---|---|---|---|---|---|
| | Near match | Wrong mapping | Different value | Missed value | External gap | Unexpected copy | Unexpected fill |
| Ontology-constrained | 0 | 3 | 21 | 13 | 59 | 0 | 7 |
| Non-ontology-constrained | 23 | 43 | 8 | 216 | 16 | 48 | 4 |
| **Total** | 23 | 46 | 29 | 229 | 75 | 48 | 11 |

*F. Evaluation excluding non-compliant reference fields*

Disagreement with an expert-curated reference record does not necessarily indicate a standardization error. Some reference values retain legacy representations that violate the target CEDAR template. A method that reproduces these values can receive credit in the original evaluation, whereas a method that follows the template may disagree with the reference. We therefore performed a second evaluation to examine how excluding fields with objectively verifiable reference violations affects the comparison between the baseline and ARMS.

We checked all reference records against the target CEDAR templates for violations of datatypes, regular-expression patterns, and closed permissible-value lists. Blank or unknown reference values did not count as violations, including for required fields. If any populated reference value violated one of these checked constraints, we excluded that field from every record in its assay. We applied exclusion at the assay-field level rather than selectively removing individual violating values to avoid evaluating a field against a partially modified reference standard. Consequently, this conservative rule also excluded record instances in which the reference value was valid. These checks identified specific, objectively verifiable template violations without subjectively reassessing the curated records. Table F.1 lists the reference fields excluded for each assay.

**Table F.1. Reference fields excluded by assay**. The audit excluded 33 assay–field combinations: 16 ontology-constrained and 17 non-ontology-constrained. Together, these exclusions removed 2,725 field comparisons across 839 records. All excluded ontology-constrained fields violated closed permissible-value lists. Among non-ontology-constrained fields, `parent_sample_id` and `data_path` violated regular-expression patterns, while `lc_gradient_value` violated its declared datatype.

| Assay name | Excluded ontology-constrained fields | Excluded non-ontology-constrained fields |
|---|---|---|
| ATACseq | `library_preparation_kit; preparation_instrument_kit; sample_indexing_kit; sequencing_reagent_kit; transposition_method; transposition_reagent_kit` | `parent_sample_id` |
| RNAseq | `barcode_read; library_preparation_kit; preparation_instrument_kit; sequencing_reagent_kit` | `parent_sample_id` |
| Auto-fluorescence | – | `parent_sample_id; data_path` |
| Cell DIVE | – | `parent_sample_id` |
| CODEX | `preparation_instrument_model` | `parent_sample_id` |
| Histology | – | `parent_sample_id; data_path` |
| IMC-2D | – | `parent_sample_id` |
| Lightsheet | – | `parent_sample_id` |
| MIBI | – | `parent_sample_id` |
| DESI | `ms_scan_mode; preparation_matrix` | `parent_sample_id` |
| LC-MS | `ms_scan_mode` | `parent_sample_id; data_path; lc_gradient_value` |
| MALDI | `ms_scan_mode; acquisition_instrument_model` | `parent_sample_id; data_path` |

Table F.2 reports the performance of ARMS and the baseline after excluding the reference fields. For ARMS, the larger changes occurred in ontology-constrained fields, where precision increased by 4.8 points and recall by 8.6 points. In non-ontology-constrained fields, precision changed little, while recall increased by 2.0 points. In contrast, the baseline showed smaller increases in ontology-constrained fields and decreases in non-ontology-constrained fields, resulting in lower overall precision and recall after exclusion. Removing fields can eliminate correct matches as well as disagreements, so exclusion does not necessarily increase performance. For example, a method can receive credit for reproducing a legacy reference value that violates the target template, but that credit is removed when the corresponding field is excluded.

**Table F.2. Field-level precision and recall before and after exclusion.** Each field in each record contributes separately to the pooled scores. Values are percentages, shown as before → after exclusion. Δ is the change from before to after exclusion for the same method, in percentage points.

| Field category | Metric | ARMS *before → after (%)* | ARMS *Δ (points)* | Baseline *before → after (%)* | Baseline *Δ (points)* |
|---|---|---|---|---|---|
| Ontology-constr. | Precision | 91.9 → 96.7 | +4.8 | 36.5 → 37.7 | +1.2 |
| Ontology-constr. | Recall | 79.9 → 88.5 | +8.6 | 33.1 → 35.1 | +2.0 |
| Non-ontology-constr. | Precision | 94.2 → 94.1 | −0.1 | 70.4 → 67.1 | −3.3 |
| Non-ontology-constr. | Recall | 87.8 → 89.8 | +2.0 | 62.6 → 59.4 | −3.2 |

We also repeated the comparison with deduplicated scoring to examine performance across distinct values rather than their frequency in the corpus. Table F.3 reports the deduplicated results before and after field exclusion. The deduplicated comparison showed a larger change in ARMS recall than the field-level comparison. ARMS recall increased by 17.3 percentage points for ontology-constrained fields and 10.6 percentage points for non-ontology-constrained fields. Before exclusion, baseline recall on distinct non-ontology-constrained values slightly exceeded ARMS recall, at 84.8% versus 84.3%. After exclusion, ARMS recall was higher, at 94.9% versus 83.3%. Precision changed less than recall under deduplicated scoring. ARMS precision increased by 4.9 percentage points for ontology-constrained fields but decreased by 0.3 percentage points for non-ontology-constrained fields.

**Table F.3. Deduplicated precision and recall before and after exclusion.** Scores are percentages, shown as before → after exclusion; Δ is the change from before to after exclusion. Recall items are distinct reference field–value pairs within assays. Precision items are assay–field combinations contributing to the paired comparison.

| Field category | Metric | ARMS *before → after (%)* | ARMS *Δ (points)* | Baseline *before → after (%)* | Baseline *Δ (points)* |
|---|---|---|---|---|---|
| Ontology-constr. | Precision | 87.8 → 92.7 | +4.9 | 32.2 → 33.5 | +1.3 |
| Ontology-constr. | Recall | 67.4 → 84.7 | +17.3 | 29.7 → 33.8 | +4.1 |
| Non-ontology-constr. | Precision | 86.0 → 85.7 | −0.3 | 68.1 → 65.5 | −2.6 |
| Non-ontology-constr. | Recall | 84.3 → 94.9 | +10.6 | 84.8 → 83.3 | −1.5 |

After excluding fields with non-compliant reference values, ARMS achieved higher deduplicated precision and recall than the baseline in both field categories. These results suggest that reference violations obscured part of ARMS's advantage in the original evaluation, and that reevaluation provides a clearer comparison of the methods on the retained fields.